\documentclass[prl,twocolumn, amsmath,amssymb,aps]{revtex4-1}
\usepackage[latin9]{inputenc}
\usepackage{bm}
\usepackage{amsmath}
\usepackage{graphicx}

\DeclareMathOperator*{\argmin}{arg\,min}
\makeatletter
\usepackage{graphicx}
\usepackage{dcolumn}
\usepackage{bm}

\usepackage[]{color}

\makeatother

\begin{document}

\preprint{APS/123-QED}
\title{Kernel methods for interpretable machine learning of order parameters}

\author{Pedro Ponte and Roger G. Melko }
\address{
	Department of Physics and Astronomy, University of Waterloo, Ontario N2L 3G1, Canada\\
	Perimeter Institute of Theoretical Physics, Waterloo, Ontario N2L 2Y5, Canada
}

\begin{abstract} Machine learning is capable of discriminating phases of matter, and finding associated phase transitions, directly from
large data sets of raw state configurations.
In the context of condensed matter physics, most progress in the field of supervised learning has come from employing neural networks
as classifiers.  Although very powerful, such algorithms suffer from a lack of interpretability, which is usually desired in scientific applications
in order to associate learned features with physical phenomena.
In this paper, we explore support vector machines (SVMs) which are a class of supervised kernel methods that provide interpretable decision functions. We find that SVMs can learn the mathematical form of physical discriminators, such as order parameters and Hamiltonian constraints, for a set of two-dimensional spin models: the ferromagnetic Ising model, a conserved-order-parameter Ising model, and the Ising gauge theory.
The ability of SVMs to provide interpretable classification highlights their potential for automating feature detection in both synthetic and experimental
data sets for condensed matter and other many-body systems.
\end{abstract}
\maketitle

\section{\label{sec:level1} Introduction}

Pattern recognition, the task of automatic discovery of features and regularities in data,
is a major focus of machine learning, which now powers many different tools in our daily lives \cite{Bishop:2006:PRM:1162264,Murphy:2012:MLP:2380985}.
From a physics perspective, the problem of searching for patterns in experimental data has long driven
theoretical progress which, by definition, relates the patterns to underlying postulates of our physical theories.
Hence, in the modern push to automate the discovery of important and novel physical features in data \cite{li2015understanding,Carrasquilla2017,leiwang,Carleo602,PhysRevB.94.165134,broecker2016machine,ch2016machine,tanaka2016detection,ohtsuki2016deep,PhysRevB.94.245129,deng2016exact,huang2017accelerated,liu2017self,torlai2016neural,vanNieuwenburg2017,zhang2016triangular,portman2016sampling,liu2017self,crawford2016quantum,aoki2016restricted,deng2017quantum,gao2017efficient,chen2017equivalence,huang2017neural,wetzel2017unsupervised,torlai2017many,hu2017discovering,schindler2017probing},
physicists must be mindful of the interpretability of machine learning results if they are
truly meant to drive theoretical process.

In condensed matter, physicists face the ultimate big-data challenge.  One must search practical
measurements, obtained from the exponentially-large state space of a system or model,
for patterns which relate to underlying theoretical paradigms.
The recent success of neural networks in classifying phases of matter \cite{Carrasquilla2017, broecker2016machine, zhang2016triangular,  PhysRevB.94.165134, Carleo602, ch2016machine,schindler2017probing} has been encouraging, in that it demonstrates how relatively
standard supervised learning tools can be repurposed for calculations in condensed matter physics.
However, contrary to industry applications of machine learning, where
performance is the prime metric of success, in physics it is generally desirable to further tie the
outcome to some theoretical structure, which can eventually be used e.g.~to make predictions.
Neural network behavior can indeed be interpreted on the simplest models of statistical mechanics,
such as the demonstration in Ref.~\cite{Carrasquilla2017} that the magnetization order parameter of the
two-dimensional (2d) Ising model is learned by the weights of the hidden units.
However, in general for more complicated models, similar success in relating network structure
to non-linear or non-local order parameters is challenging, especially in the case of deep neural networks.
This lack of interpretability presents a challenge for the goal of driving theoretical progress with
machine learning.  It is therefore crucial that the condensed matter community  survey the
performance of interpretable machine learning algorithms on data obtained from models of
interest to condensed matter physics.

In this paper we study interpretable supervised learning
algorithms applied to the discrimination of phases of matter
in large synthetic data sets produced by numerical simulation.
We focus on a particular class
of machine learning algorithms called support vector machines (SVMs).
In particular, we ask whether SVMs in the supervised learning setting
are able to discover the mathematical structure of physical order parameters.
We first introduce the SVM algorithm and describe its general properties,
with particular focus on the {\it kernel trick}.
This trick allows us to perform linear regression on non-linear
features of the data, without explicitly generating the features.
In particular, this is relevant when applied to Monte Carlo data
from many-body systems, which typically have high-dimensional
physical states. Without the kernel trick, it would be unpractical
to generate a set of non-linear features, which could be
exponentially large.
In section \ref{sec:Results}, we perform phase classification
on Monte Carlo configurations produced from the 2d Ising model, a conserved-order-parameter
Ising model (COP) \cite{leiwang}, and the 2d Ising gauge theory.
We find that SVMs with a quadratic polynomial kernel can discriminate the phases of the 2d Ising model by
learning the correct physical order parameter (the squared magnetization per spin).
In the case of COP model, we find that SVMs are able to discover the non-trivial order parameter,
devised by Wang \cite{leiwang} through visualization of dimensionally-reduced data,
with less human intervention.
Finally, we show that the 2d Ising gauge theory, whose $T=0$ ground state is
defined by local plaquette constraints but no conventional order parameter,
can be discriminated by SVMs with polynomial kernels.
In this case, no polynomial with
order less than four can identify the ground state,
indicating that the SVM is able to learn the original Hamiltonian in order to evaluate whether
the local constraints on four Ising degrees of freedom are satisfied for each plaquette.

\section{Support Vector Machines \label{sec:Support-Vector-Machines}}

Linear support vector machines \cite{Murphy:2012:MLP:2380985} were initially designed
to perform binary classification. They belong to the class of supervised
learning algorithms that, given an input $\bm{x}\in\mathcal{\mathbb{\mathbb{R}}}^{p}$,
predict the class $y\in\{-1,1\}$ in which it is most likely to belong. In the context of this paper, the input $\bm{x}$ will represent a spin configuration from a statistical mechanical model of interest in condensed matter physics, in which the different components of $\bm{x}$, called {\it features} in the machine learning literature, correspond to the spins at different lattice sites. The output $y$ will label the thermodynamic phase it was sampled from.
The main idea behind SVMs is to find the hyperplane,
defined by $\bm{w}\cdot\bm{x}+b=0$, that best separates the two classes. Formally, this is achieved by solving the optimization problem:
\begin{align}
 & \argmin_{\bm{w},\xi_{i},b} \left\{ \frac{1}{2}\bm{w}\cdot\bm{w}+C\sum_{i=1}^{N}\xi_{i} \right\}\label{eq:svm_optimization} \nonumber \\
 & \mathrm{such\ that\ }y^{(i)}(\bm{w}\cdot\bm{x}^{(i)}+b)\geq1-\xi_{i},\;i=1\ldots N.
\end{align}
Here, $C$ is a constant, $\xi$ are ``slack'' variables (described more below) and
 $N$ is the number of input samples, called the training set.
Finding optimal parameters $\bm{w}$ and $b$, which can then be used to make predictions
on a test set, is the goal of supervised learning with SVMs.
 Unlike say a feed-forward neural network \cite{Carrasquilla2017}, this model does not provide
probabilistic predictions. However, the optimization problem Eq.~\eqref{eq:svm_optimization} is equivalent to
$\argmin_{\bm{w},b}\left\{\frac{1}{2}\bm{w}\cdot\bm{w}+C\sum_{i=1}^{N}\max\left(0,1-y_{i}d(\bm{x}^{(i)})\right) \right\}$
with $d(\bm{x}^{(i)})=\bm{w}\cdot\bm{x}^{(i)}+b$. Because the hinge
function $\max(0, 1 - yd)$ approximates the misclassification error $\Theta(-yd)$ \cite{Bishop:2006:PRM:1162264}, this can be viewed as an approximation
to minimizing the misclassification error with
a so-called $\ell_{2}$-norm regularization (which penalizes unnecessary coefficients of $\bm{w}$).
This analogy also explicitly shows that $C$ can be interpreted as a regularization
parameter.

When the $\xi_i$ variables are constrained to be zero in Eq.~(\ref{eq:svm_optimization}),
this corresponds to the hard margin case and the optimization algorithm has a solution only for linearly
separable classes. In this case, the solution corresponds to the hyperplane
$\bm{w}\cdot\bm{x}+b=0$ for which the {\it margin}, defined as the minimum
distance $d_{\mathrm{min}}=1/||\bm{w}||$ of the data samples to the
hyperplane, is maximum. Thus, this algorithm finds the hyperplane
which maximizes the training set margin and provides the most confident
predictions on new inputs. On the other hand, perfectly separable
data is not a typical property of datasets and in general the constraints
$y^{(i)}(\bm{w}\cdot\bm{x}^{(i)}+b)\geq1$ are not feasible. Hence,
it is necessary to introduce the slack variables $\xi$ which allow a
training set input $\bm{x}^{(i)}$ to violate the margin at a cost
of $C\xi_{i}$. For a given training sample, the slack variable $\xi_{i}$ can take different values.
 $\xi_{i}=0$ if the margin is not violated;  $0 < \xi_{i} \le 1$ if it is on the correct side of the hyperplane
but violates the margin;
and  $\xi_{i}>1$ if the sample is misclassified. In general it is
necessary to use a test set to find the optimal parameter $C$. This
will provide the best trade-off between minimizing training errors
and the model complexity.

After optimization (i.e.~training), the class to which a new input $\bm{x}$ belongs
is predicted as $y=\mathrm{sign}(d(\bm{x}))$, hence $d(\bm{x})$ is
referred to as the
{\it decision function} in the machine learning literature. Even though
the previous optimization problem can be solved by quadratic programming \cite{Murphy:2012:MLP:2380985},
it also admits a dual formulation whereby the primal variables $\bm{w}$,
$\xi$, $b$ are eliminated and the optimization is performed over
$N$ dual variables $\alpha_{i}$,
which are the Lagrange multipliers associated
with each constraint in Eq.~(\ref{eq:svm_optimization}).
The optimal parameter $\bm{w}$ is then expressed as
$\bm{w}=\sum_{i=1}^{N}\alpha_{i}y_{i}\bm{x}^{(i)}$.
A crucial feature of this dual formulation is that the optimization algorithm only depends
on inner products of the training samples $\langle\bm{x}^{(i)},\bm{x}^{(j)}\rangle$.
In addition, at prediction time, one only needs to calculate the inner product between the
training samples and new samples. Because the algorithm is formulated such that the
input vector enters only in the form of a scalar product,
this allows us to employ the {\it kernel trick}, whereby we replace $\bm{x}^{(i)}\cdot\bm{x}^{(j)}$
with some other choice of kernel function, $K(\bm{x}^{(i)},\bm{x}^{(j)})$.
Then, at prediction time the decision function has the form
\begin{equation}
d(\bm{x}) = \sum_{i=1}^N \alpha_i y_i K(\bm{x}^{(i)}, \bm{x}) + b, \label{GeneralDF}
\end{equation}
whereby we can learn more complex decision functions depending on
the choice of the kernel without explicitly generating more
features in our input $\bm{x}$.
For example, the
kernel $K(\bm{x}^{(i)},\bm{x}^{(j)})=(\bm{x}^{(i)}\cdot\bm{x}^{(j)}+c_{0})^{d}$
corresponds to the mapping to a ${{p+d} \choose d}$ dimensional feature
space corresponding to all the monomials of the form $x_{i_{1}}x_{i_{2}}\ldots x_{i_{k}}$
(ignoring permutations) that are up to order $d$ where $p$ is the
number of \textit{raw} features.

\section{Results}\label{sec:Results}

In this section, we perform supervised learning with SVMs on data sets generated by sampling spin configurations of classical
Hamiltonians, where Ising degrees of freedom $\bm{\sigma}$ (``spins'' taking binary values) will serve as our input $\bm{x}$. Finite size lattices with $N$ spins are considered.
In the below, we use the most efficient training algorithm for SVMs by means of the scikit-learn library \cite{scikit-learn} -- the Sequential Minimal Optimization algorithm \cite{platt} -- known to scale as $O(N^{2})$, or with a smaller power, for several kernels and types of data. We find, in practice, training on a single core
is generally slow for $10^{5}$ samples or more. We explore the behaviour of different polynomial kernels and perform grid search to find the optimal regularization parameter $C$. In general, the results are averaged over several choices of training and test sets for the same hyperparameters in order to obtain more reliable statistics.

\subsection{2d Ising Model\label{subsec:2d-Ising-Model}}

We first consider the nearest-neighbor Ising model in two dimensions, $H = - \sum_{\langle \bm{a} \bm{b} \rangle} \sigma_{\bm{a}} \sigma_{\bm{b}}$, where $\sigma_{\bm{a}}=\pm1$, and $\bm{a}$ are the euclidean coordinates of a given lattice site.
Monte Carlo simulations using the Wolff algorithm were performed to
collect spin configurations $\bm{\sigma}^{(i)}=(\ldots,\sigma_{\bm{a}}^{(i)},\ldots)$
where $i$ identifies each configuration
collected at different temperatures from $T=1.6$
to $T=2.9$. The 2d Ising phase transition occurs at the critical
temperature $T_{c}\approx 2.269$ \cite{PhysRev.65.117} and separates
a ferromagnetic (FM) phase, characterized by a non-zero total magnetization
per spin, from a featureless paramagnetic (PM) phase at high temperatures. For a given $L \times L = N$ size lattice,
each sample is labeled with its corresponding phase in the binary class $y_{i}=\pm1$.
We train SVMs to learn to discriminate between the two phases for
different numbers of samples in the training set. For the Ising model, we limit our survey to a linear
and a quadratic kernel of general form $K(\bm{\sigma},\bm{\sigma}^{\prime})=(\bm{\sigma}\cdot\bm{\sigma}^{\prime}+c_{0})^{k}$
with $k=1,2$ and $c_{0}=0$. Note that in general, it might be necessary
for the learning procedure to find the optimal $c_{0}$ as well.

\begin{figure}
\includegraphics[width=0.9\columnwidth]{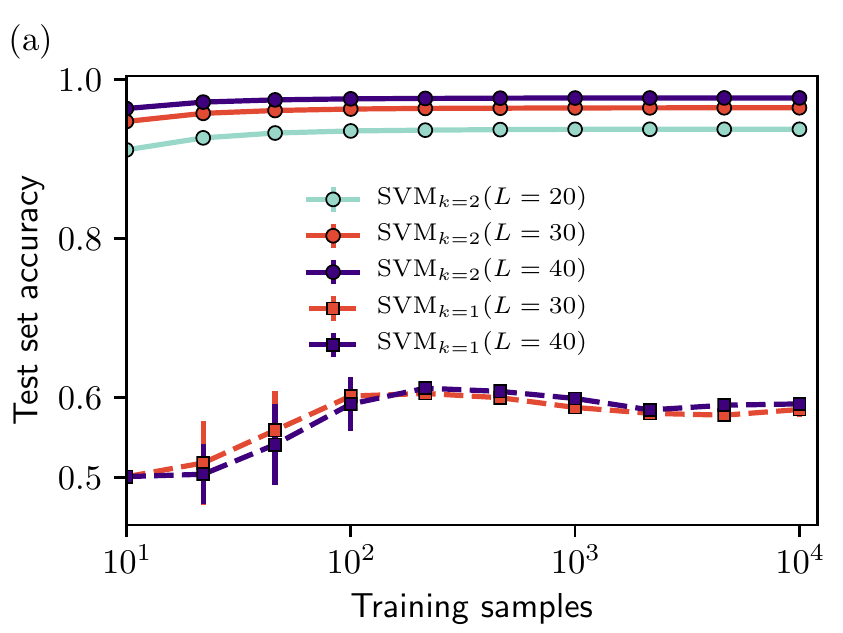}
\includegraphics[width=0.9\columnwidth]{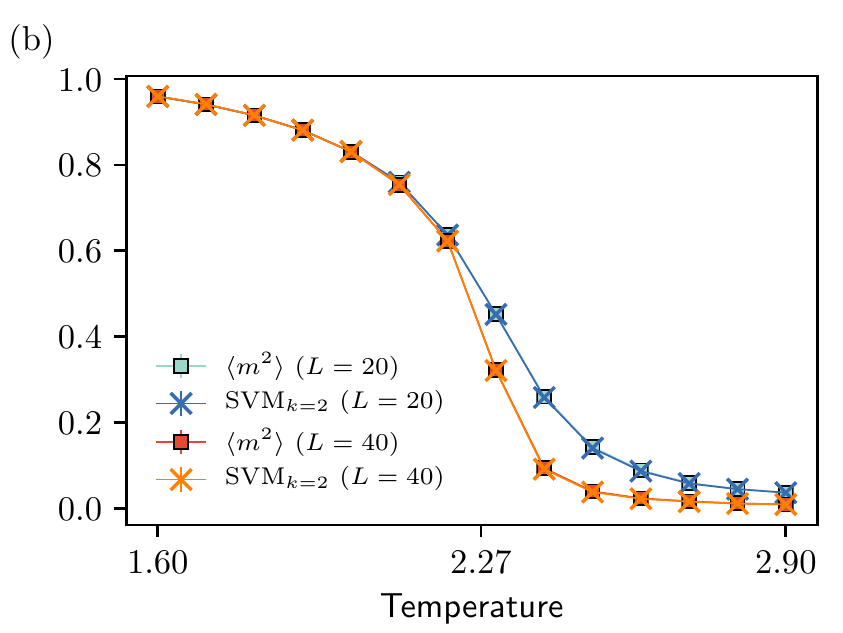}
\caption{\label{fig:ising_accuracy_comparison} (a) Average test
set accuracy of Support Vector Machines with polynomial kernel $K(\bm{\sigma},\bm{\sigma}^{\prime})=(\bm{\sigma}\cdot\bm{\sigma}^{\prime})^{k}$
trained on Monte Carlo sampled configurations from the 2d Ising model.
For each number of training samples, the accuracy is averaged over
100 independent training and test sets. (b) The SVM classifies
samples according to $\mathrm{sign}(d(\bm{\sigma}))$. The decision
function $d(\bm{\sigma})$ for the SVM with a quadratic polynomial kernel is evaluated
by Monte Carlo sampling at different temperatures and compared to
the squared magnetization per spin $m^{2}$. The arbitrary scale factor
and off-set in the SVM decision function are fixed by matching
the decision function to $\langle m^2 \rangle$
at $T=1.6$ and $T=2.9$.
}
\end{figure}

\begin{figure}
\includegraphics[width=0.45\columnwidth]{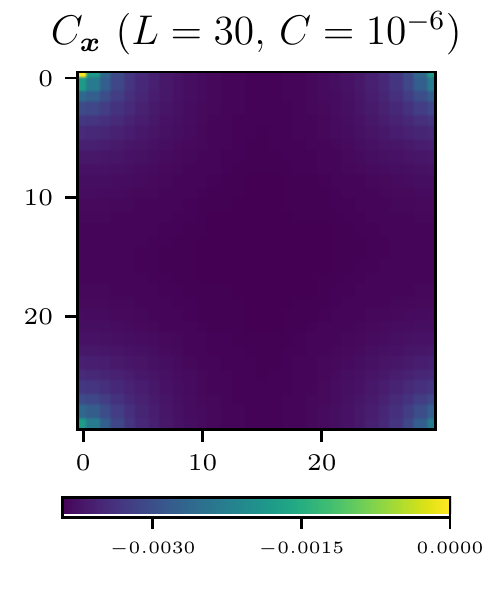}
\includegraphics[width=0.45\columnwidth]{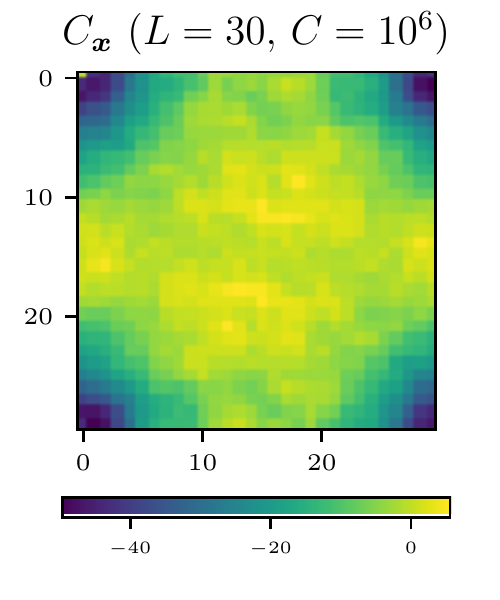}
\caption{ \label{fig:ising_op_heatmap}The decision function for an SVM with
quadratic polynomial kernel, Eq.~\eqref{eq:decision_function}.
The averaged $C_{\bm{x}}=\langle C_{\bm{x}}^{(\bm{a})}\rangle_{\bm{a}}$
is displayed for different values of regularization $C=10^{-6}$ and
$C=10^{6}$. Clearly, for large regularization ($C=10^{-6}$), the
decision function is essentially the $m^{2}$ order parameter of the
2d Ising model.}
\end{figure}

In order to quantify the performance of each SVM model, the main metric that we
study is the test set accuracy as a function of the number
of samples in the training set, for the value of $C$ which results
in the optimal accuracy of the model, illustrated in
Fig.~\ref{fig:ising_accuracy_comparison} (a).
For the linear kernel, the exploration of $C$ is over a log-spaced
grid of 11 values from $10^{-5}$ to $10^{5}$.
For the quadratic
kernel the accuracy does not depend significantly on the choice of
$C$ and thus we fixed it at $C=10^{-5}$. For a given number of
training samples and regularization $C$, the test set accuracy is
additionally averaged over different random selections of training
and test sets.

Results for the test set accuracy and for the SVM decision function
are shown in Fig.~\ref{fig:ising_accuracy_comparison}. As seen in Fig.~\ref{fig:ising_accuracy_comparison} (a), the quadratic kernel performs extremely well with mean
test set accuracy $\approx97\%$ for $L=40.$ This can be easily {\it interpreted}, since
we know that this model possesses a quadratic order parameter that linearly
discriminates the FM from the PM,
i.e.~the squared magnetization per spin $m^{2}=(\sum_{\bm{a}}\sigma_{\bm{a}}/N)^{2}$.
We find that the quadratic kernel reaches very significant
performance with only a few dozen samples in the training set, which
is a result of the simplicity of this model. Moreover, with
increasing number of samples the test set accuracy approaches a plateau
value which increases with system size towards $100\%$. This
is the expected behaviour, since at the critical point the fluctuations
of the order parameter approach zero in the thermodynamic limit and
it is thus possible to discriminate perfectly between both phases.

For the linear kernel (Fig.~\ref{fig:ising_accuracy_comparison} (a)),
the accuracy shows non-monotonic behaviour
with the total number of training set samples and does not improve
with increasing system size. This is a consequence of the fact that a
linear decision function is unable to discriminate between the FM and PM phases.
Namely, in the FM phase configurations have magnetization per spin near $\pm1$,
while for the PM phase most configurations have appproximately zero magnetization.
Thus, the $k=1$ kernel is asking a linear decision boundary to separate a data
set with three clusters -- an impossible task.
Close inspection of the decision function learned by the
SVM reveals it contains random linear coefficients without any structure,
confirming that nothing physically relevant is being learned about the data in this case.

As noted above, the accuracy of the SVM with a quadratic kernel
on the test set does not depend significantly on the regularization
parameter $C$. An advantage of SVMs is that we can visualize the
decision function being learned. From Eq.~\eqref{GeneralDF}, the decision function
for an SVM with quadratic polynomial kernel can be expressed as
\begin{align} d(\bm{\sigma})=\sum_{\bm{a}}\sum_{\bm{x}}C_{\bm{x}}^{(\bm{a})}\bm{\sigma}_{\bm{a}}\bm{\sigma}_{\bm{a}+\bm{x}}+b. \label{eq:decision_function}
\end{align}
In Fig.~\ref{fig:ising_op_heatmap}, we display the heatmap of $C_{\bm{x}}=\langle C_{\bm{x}}^{(\bm{a})}\rangle_{\bm{a}}$
, where $\langle\ldots\rangle_{\bm{a}}$ denotes averaging with respect
to all sites $\bm{a}$ for $C=10^{-6}$ and $C=10^{6}$ and system size
$L=30$. It is interesting to note that even though the classification performance
is very similar, the SVM decision function corresponds to different order parameters depending
on the amount of regularization. Clearly, at $C=10^{-6}$, the SVM
is learning $m^{2}$ as the order parameter of the model up to finite-size effects.
In contrast, at $C=10^{6}$, the
SVM is learning to calculate the square of the total magnetization
within some fixed distance of each spin and summing all these different
local contributions. To further illustrate this point, in Fig. \ref{fig:ising_accuracy_comparison} (b), the SVM decision
function (with $C=10^{-6}$ regularization) is averaged
over Monte Carlo samples at different temperatures showing essentially perfect
agreement with $m^{2}$. Of course, the SVM decision function has an
arbitrary scale and off-set and in order to match with $\langle m^{2}\rangle$ a linear
transformation is performed, so that they agree at the extreme values
of temperature $T=1.6$ and $2.9$.

\begin{figure}
\includegraphics[width=0.9\columnwidth]{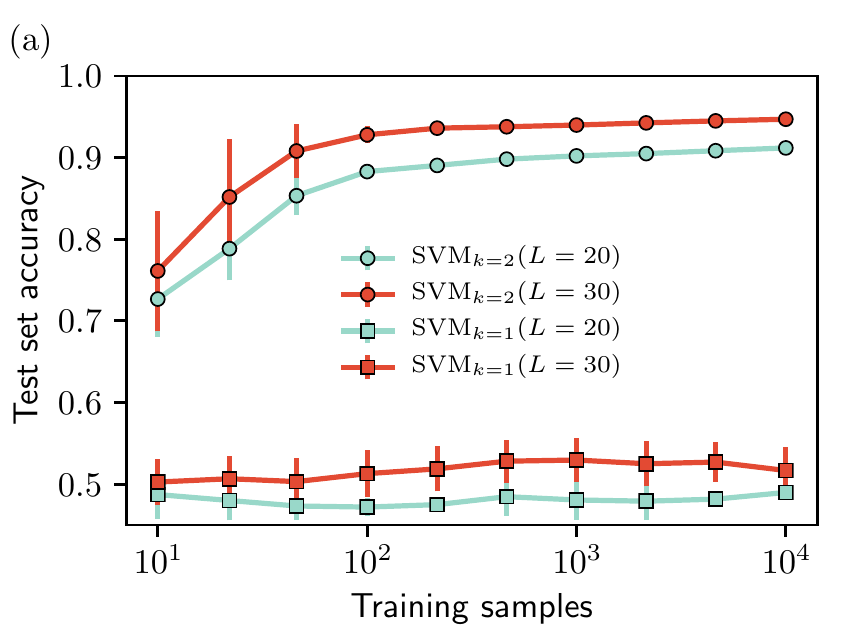}
\includegraphics[width=0.9\columnwidth]{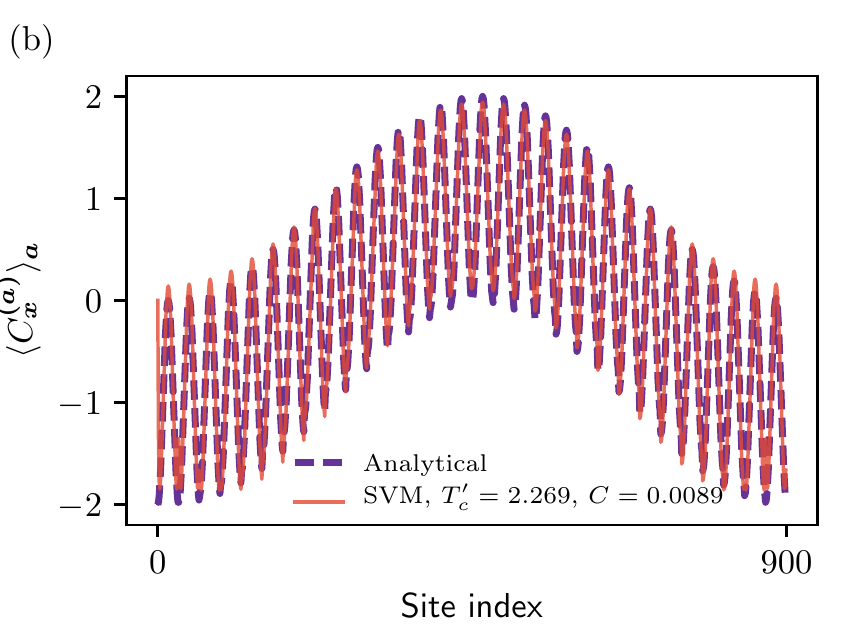}

\caption{\label{fig:COP_accuracy_spatial_dependence}(a) Average test set accuracy vs number of training samples for Support
Vector Machines with polynomial kernel $K(\bm{\sigma},\bm{\sigma}^{\prime})=(\bm{\sigma}\cdot\bm{\sigma}^{\prime})^{k}$
trained on Monte Carlo sampled configurations from the conserved-order-parameter
Ising model at different temperatures. 
(b) The spatial dependence of the SVM decision function coefficients $\langle C^{(\bm{a})}_{\bm{x}} \rangle_{\bm{a}}$ learned by an SVM
with quadratic kernel and regularization coefficient
$C=0.0089$ for system size $L=30$ shows very good agreement with the analytical form (\ref{Eq:WangOrderParameterCoefficients}) devised in \cite{leiwang}.}

\end{figure}

\begin{figure}
\includegraphics[width=0.9\columnwidth]{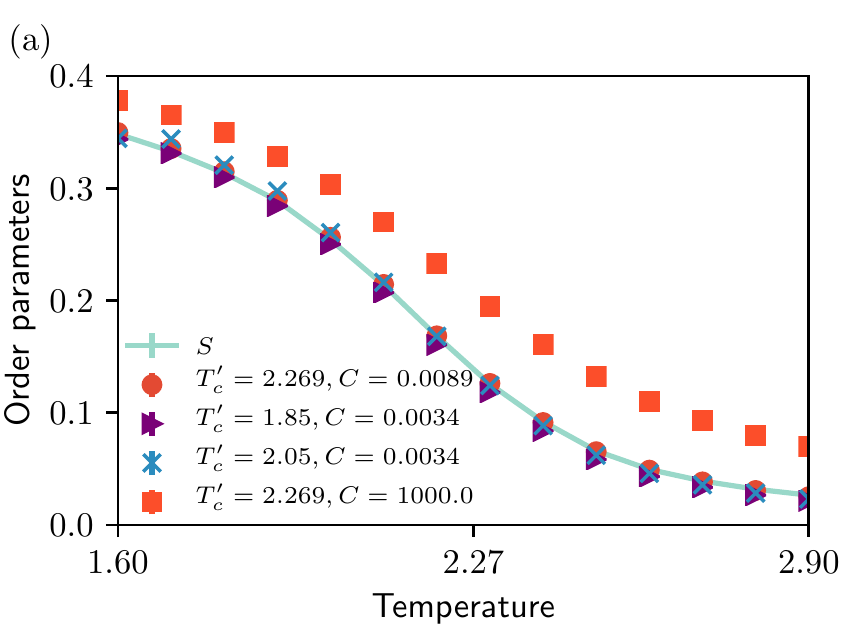}
\includegraphics[width=0.9\columnwidth]{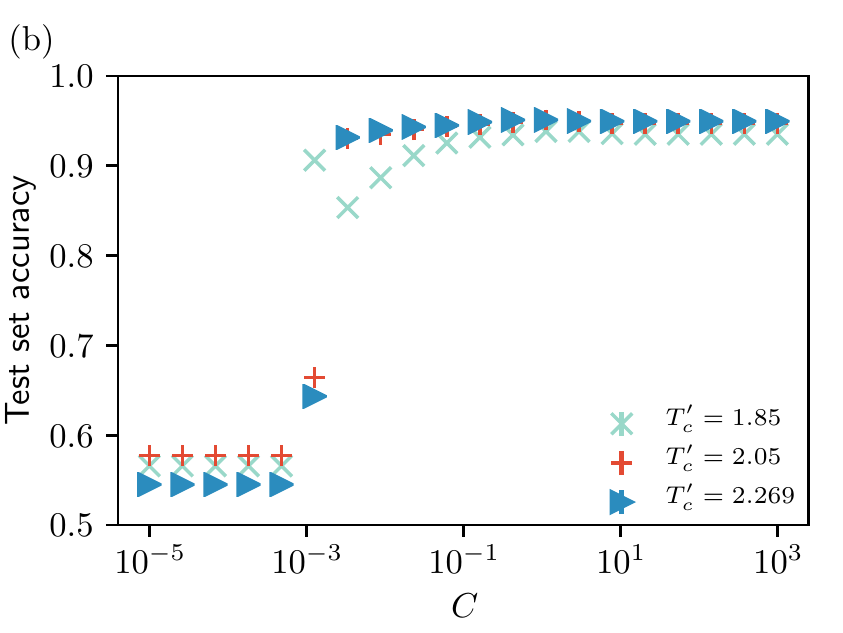}

\caption{\label{fig:COP_different_Tc_regularization}(a) Comparing the decision function averaged over Monte Carlo samples at different temperatures learned by SVMs
with quadratic kernel - trained on $30000$ samples with
$L=30$ - to the analytical order parameter $S$ (Eq. (\ref{Eq:WangOrderParameter})) assuming
different critical temperatures $T_{c}^{\prime}$ and amount of regularization
$C$. (b) Dependence on regularization $C$
of the test set accuracy (averaged over 10 sets of $10000$ samples)
of quadratic SVMs trained on $30000$ samples for $L=30$.
}
\end{figure}

\subsection{Conserved-order-parameter Ising model\label{subsec:Conserved-order-parameter-Ising-}}

The conserved-order-parameter Ising model \cite{leiwang} is described by the same
Hamiltonian as the Ising model but the configuration space is restricted
to the subspace where the total magnetization $\sum_{\bm{a}}\sigma_{\bm{a}}$
is zero. This model describes a half-filled lattice gas of particles
with a nearest neighbour attractive interaction \cite{newman1999monte}.
At low temperatures,
domains of up and down spins are separated by either two horizontal
or two vertical domain walls for a square lattice geometry with periodic
boundary conditions. There is a phase transition to a featureless
phase at the same critical temperature as the 2d Ising model. Ref.
\cite{leiwang} studied this model on the square lattice using Principal Component Analysis (PCA),
 a dimensional reduction algorithm, and devised the following order parameter:
\begin{align}
S=&\frac{1}{L^{4}}\sum_{\bm{a},\bm{b}}\sigma_{\bm{a}}\sigma_{\bm{b}}\left[\cos\left(\frac{2\pi}{L}(a_1-b_1)\right)\right. \nonumber\\
&\left.+ \cos\left(\frac{2\pi}{L}(a_2-b_2)\right)\right] . \label{Eq:WangOrderParameter}
\end{align}
$S$ is of the form of Eq.~(\ref{eq:decision_function}) with
\begin{align}
C^{(\bm{a})}_{\bm{x}} = \frac{1}{L^4}\left[\cos\left(\frac{2\pi}{L}x_1\right)+  \cos\left(\frac{2\pi}{L}x_2\right)\right].
\label{Eq:WangOrderParameterCoefficients}
\end{align}
In contrast to PCA, which through feature dimension reduction provides the easy visualization required for the determination
of this order parameter, SVMs should provide automated order parameter detection in a more systematic way.  We explore this
ability now.

As in the case with the Ising model, we collect Monte Carlo samples
for a set of temperatures below and above $T_{c}$. Fig. \ref{fig:COP_accuracy_spatial_dependence}
(a) shows the averaged test set accuracy as a function
of the number of samples for the optimal value of the
regularization parameter $C$ over a grid of 11 values from $10^{-5}$ to $10^5$.
The results are averaged over $400$
training and test sets. As in the Ising model case, for a quadratic
kernel there is a monotonic improvement of test accuracy with the
number of training samples.  The limiting value as the number of
training samples becomes very large
also increases with system size. This behavior signals
the existence of a quadratic order parameter that discriminates between
the two phases. As a check, we compare the
explicit spatial dependence $C_{\bm{x}}=\langle C_{\bm{x}}^{(\bm{a})}\rangle_{\bm{a}}$ of the SVM decision function for $C=0.0089$ (the selection of this regularization value is discussed below) to the analytical form (\ref{Eq:WangOrderParameterCoefficients}). The scale and off-set of the SVM decision function are fixed by requiring the coefficients $C_{\bm{x}}$ to have the same mean and standard deviation as the analytical form.
Fig. \ref{fig:COP_accuracy_spatial_dependence} (b) shows very good agreement
between them. In contrast, the linear kernel does not show
a clear improvement of the test set accuracy with increasing number
of samples which indicates there isn't an order parameter of that
form.

One could ask whether this order parameter could also be learned
in the case where the precise value of the critical temperature $T_c^{\prime}$
is not known.
We address this by performing supervised learning with quadratic SVMs assuming
different values $T_{c}^{\prime}$ for the critical temperature.
Finite-size scaling of the learned order parameter (or its moments)
can then be used to estimate the value of the critical temperature.
Fig.~\ref{fig:COP_different_Tc_regularization}
(a) compares the actual order parameter $S$ (Eq. (\ref{Eq:WangOrderParameter})) with the
SVM decision function for different values of $T_c^\prime=1.85,\,2.05,\,2.269$.
The scale and off-set of the SVM decision function are fixed by matching it with $S$ at $T=0.1$
and $T=100.0$. We observe that
the SVM decision functions learned when assuming $T_{c}^{\prime}=1.85,2.05$
(for a choice of regularization values discussed below) also agree well
 with $S$, suggesting machine learning
algorithms are able to learn important physical information without
the precise knowledge of $T_{c}$.

We now analyze the role of the regularization parameter $C$ for learning in this model. Fig. \ref{fig:COP_different_Tc_regularization} (b), shows for $L=30$ and a training
set with $30000$ samples, the test set accuracy averaged over 10
randomly picked test sets of $10000$ samples versus the regularization
parameter $C$ for $T_c^\prime=1.85,\,2.05,\,2.269$.
At $C\gtrsim10^{-3}$,
the test set accuracy jumps significantly to very high values ($\approx 90\%$)
and then reaches a plateau. Interestingly, we find that throughout the plateau region
the decision function of the SVM changes continuously. Fig. \ref{fig:COP_different_Tc_regularization}
(a) also compares the analytical order parameter $S$ with the
SVM decision function for different values of $C$ when $T_c^\prime=T_c$. For $C=0.0089$, right at the beginning of the plateau,
there is highly accurate agreement between $S$ and the SVM decision function,
while for $C=1000$ the two do not match. Similar results were also found for the other values of $T_c^\prime$.
Thus, the physical order parameter can be interpreted as being
associated with the least amount of complexity that
still allows the model to have good predictive performance. This was
also what we found for the Ising model in Section \ref{subsec:2d-Ising-Model}.

\subsection{2d Ising Gauge theory\label{subsec:2d-Ising-Gauge}}

Finally, we consider the challenging case of topologically ordered systems, where
no conventional local parameter exists.
For concreteness we study the 2d Ising gauge theory with Hamiltonian $H=-\sum_{p}\prod_{i\in p}\sigma_{i}^{z}$,
where the spins live on the bonds of a 2d square lattice and $p$
represents a plaquette with four spins. The set of ground states is
a degenerate manifold with the constraint that for all plaquettes
$p$, $\prod_{i\in p}\sigma_{i}^{z}=1$. In the thermodynamic limit,
the constraints are violated at any finite temperature. A
conventional order parameter that distinguishes ground states from
finite temperature states does not exist.
Ref.~\cite{Carrasquilla2017} found that the simplest fully connected
feed forward neural networks were unable to classify these two different
cases in a supervised learning context.  Only with convolutional neural
networks, which explicitly exploit locality and translational invariance, did this
classification task succeed on raw spin configurations.
We note that by engineering predictive features as a pre-processing step,
simple feed forward neural networks may be successful in classification
(as demonstrated in Ref. \cite{zhang2016triangular}).
However in the present case, we explore the behaviour of SVMs for the raw 2d Ising Gauge theory data
with no pre-processing on the input data.

\begin{figure}
\includegraphics[width=0.9\columnwidth]{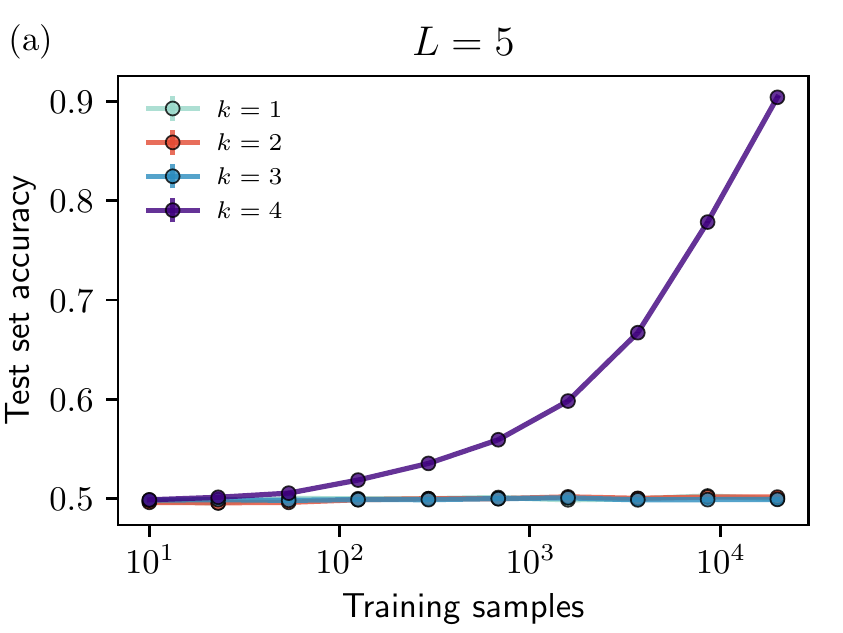}
\includegraphics[width=0.9\columnwidth]{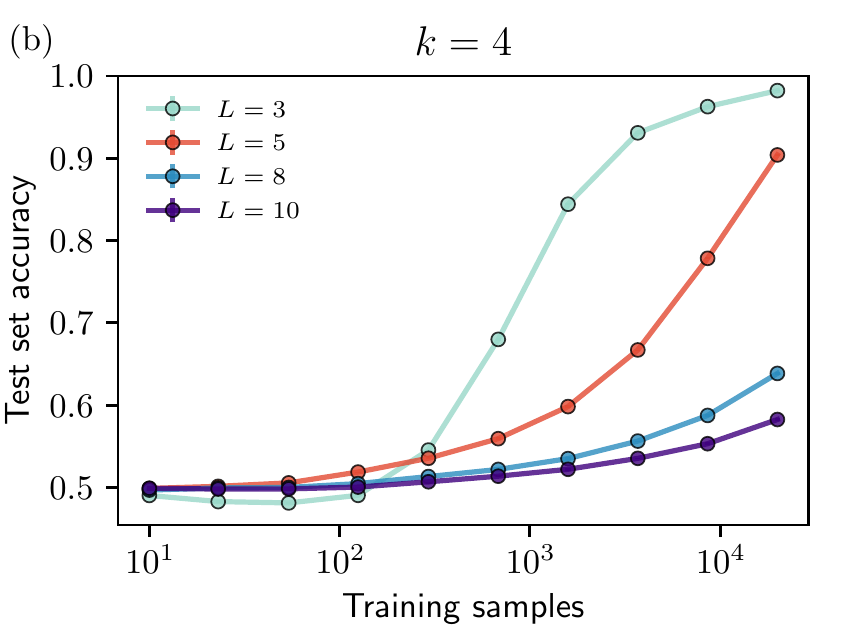}
\caption{\label{fig:toric_code} (a) Test set accuracy of SVMs with
polynomial kernels of order $k$ in classifying ground states versus
infinite temperature states of the 2d Ising Gauge theory for system
size $L=5$. Only in the feature space of fourth order polynomials is
the SVM algorithm performance better than a random algorithm. The
performance approaches 100\% with increasing number of training samples.
(b) Test set accuracy for $k=4$ and different system sizes. For
larger system sizes, more training samples are needed to learn the
correct decision function.}
\end{figure}

We perform supervised learning on spin configurations generated for the
Ising Gauge theory at $T=0$ and $T= \infty$ (i.e.~completely random spin states).
We explore kernels of increasing degree, starting from $k=1$, for different
system sizes $L$.
Fig. \ref{fig:toric_code} (a) shows that for system size
$L=5$, SVMs with polynomial kernel of degree less than 4 fail to
discriminate these two phases, exhibiting average test set accuracies
$\sim50\%$, which amounts to random guessing. However, a SVM with polynomial
kernel of fourth order is able to perform the task with an accuracy
that converges towards $100\%$ as the number of training samples
is increased.
As apparent in Fig. \ref{fig:toric_code} (b), we also note
 that the number of training samples necessary
to learn the fourth order discriminator increases with system size.
This illustrates the difficulty that SVMs can have in performing classification for large system sizes.

\begin{figure}
\centering \includegraphics[width=0.8\columnwidth]{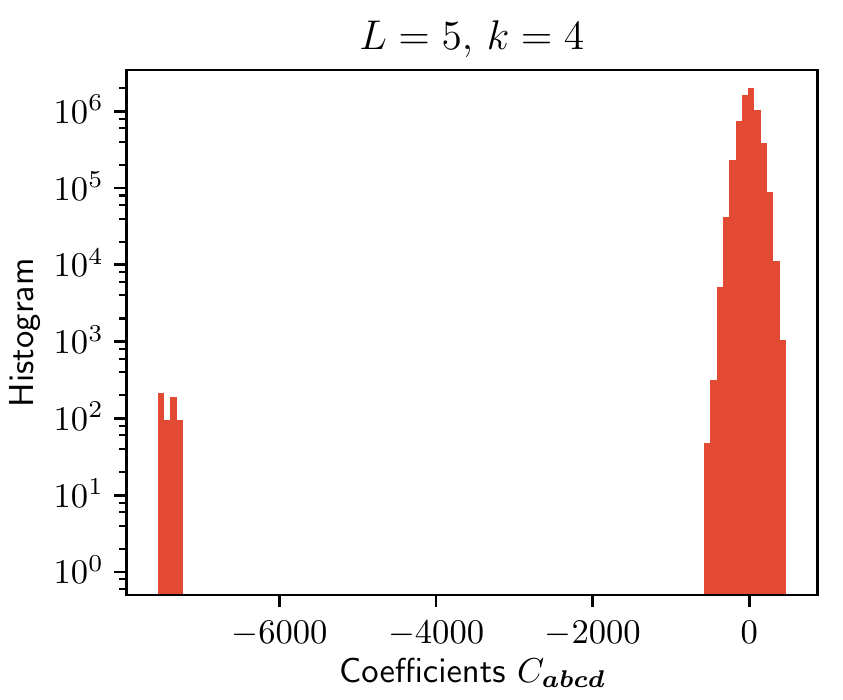}
\caption{\label{fig:toric_code_hist}Histogram of the coefficients $C_{\bm{a}\bm{b}\bm{c}\bm{d}}$
in the decision function
learned by a SVM with 4th order polynomial kernel in classifying ground
states versus infinite temperature states of the 2d Ising Gauge theory
for $L=5$. There are $600$ large coefficients (in absolute value)
which correspond to the 25 plaquettes and the permutations of their
spins.}
\end{figure}

In order to interpret these results, we further analyze the decision function of the SVM.
The smallest degree (fourth order) polymonial kernel learns a decision function of the
form $d(\bm{\sigma})=\sum_{\bm{a}\bm{b}\bm{c}\bm{d}}C_{\bm{a}\bm{b}\bm{c}\bm{d}}\sigma_{\bm{a}}\sigma_{\bm{b}}\sigma_{\bm{c}}\sigma_{\bm{d}}+b$, which contains the product of four spins.
Fig. \ref{fig:toric_code_hist} shows the histogram of the coefficients
$C_{\bm{a}\bm{b}\bm{c}\bm{d}}$ for an SVM trained on 20000 samples
and $C=10^{6}$.
From this plot it is clear that there are two sets of coefficients; one near zero, and another set with
large negative values.
Counting the number of these large (in magnitude) coefficients reveals 600.
For this lattice, with 25 square plaquettes, this number corresponds to the the number of possible
permutations of the four indices of ${\bm{a}\bm{b}\bm{c}\bm{d}}$, i.e.~$25\times 4!$.
Hence, this model provides evidence that SVMs are able to learn complex interpretable
decision functions and discover the locality of the Hamiltonian directly from raw data
on spin configurations.

\section{Conclusions}

We have examined the use of support vector machines
(SVMs), one of the most common tools for supervised learning, for the
binary classification of phases in several models of interest to
condensed matter and statistical physics.
SVMs employ a {\it kernel trick} to define a decision function, used to
discriminate features in a higher dimensional space.
The kernel depends on the inner product of spin vectors, but can otherwise have
some freedom in definition.
We have shown that in contrast to other methods such as neural
networks, the ability to use different kernels for classification tasks
gives SVMs significant value in finding {\it interpretable} physical discriminators
for different models, such as conventional order parameters defined in condensed matter theory.

To allow for some slack to misclassify data, SVMs employ a regularization, which controls the
tradeoff between minimizing training errors, while still allowing sufficient model complexity.
In this work we have found that the amount of regularization has a strong impact on the decision function learned.
We find that the expected physical order parameter is associated with the least complex model
(i.e.~the largest amount of regularization) that is still able to obtain near-optimal test set accuracy.

On the other hand, we find that small values of regularization that achieve
the same (or better) performance than the physical order parameter
learn a non-physical decision function -- one
not related to any conventional order parameter.
Since such decision functions generalize well to uncorrelated test set samples, we argue that this is not
an example of overfitting, even though it may arise due to particular details contained within the training set.
This observation deserves further study, as it may have consequences
more generally for the role of regularization in black box algorithms, such as neural networks,
when applied to data in condensed matter physics.

While very successful on the Ising-like examples studied in this paper, the SVM algorithm is not
without its limitations.
As observed for the 2d Ising Gauge Theory, the number of samples required to learn the
physically-relevant decision function can grow prohibitively large.
The reason that this occurs for the degenerate groundstate of the Ising gauge theory, and not the models
with conventional order parameters, desires further study.
We generally observe that the training of SVMs
when the number of samples is larger than $10^{5}$ becomes time-consuming.
This could also be an issue in using SVMs near phase transitions in models
where a large number of samples is necessary to learn the physical discriminator.
In such cases, it would be interesting to further explore the interpretability of neural networks,
since they can possibly present better scalability to larger system sizes.

It would be interesting to study the utility of SVMs on other classical models of
interest in condensed matter physics.  We note that for systems with continuous degrees
of freedom it might be necessary to consider a radial basis function
as a kernel, which maps to an infinite dimensional space, instead
of polynomials. For example, it would interesting to see if for the
2d XY model, the spin stiffness can be identified as the physical discriminator
for the Kosterlitz-Thouless transition using a suitably-modified SVM.
Finally, SVMs and other sparse kernel machines may in the future be
easily adapted to study quantum phases and phase transitions by using wavefunctions
or density matrices as data. SVMs with linear or quadratic
polynomial kernels could be used to determine operators which
discriminate between different quantum phases using density
matrices or wavefunctions as data.  Thus, as the condensed matter
community increasingly adopts modern machine learning methods into its numerical arsenal,
we expect that SVMs will become a standard tool for finding {\it interpretable} physical
discriminators for generic Hamiltonians in the near future.

{\it Acknowledgements --}
The authors thank Juan Carrasquilla for enlightening discussions. This research was supported by NSERC, the
Canada Research Chair program, grant SFRH/BD/84875/2012
from Funda\c{c}\~ao para a
Ci\^encia e a Tecnologia (Portugal), and
the Perimeter Institute for Theoretical Physics. Simulations were performed on resources provided by SHARCNET. Research at Perimeter Institute is supported through Industry Canada and by the Province of Ontario through the Ministry of Research \& Innovation.

\bibliography{paper.bib}
\end{document}